\documentclass[draft]{agujournal2019}
\usepackage{url} 
\usepackage{lineno}
\usepackage[inline]{trackchanges} 
\usepackage{soul}
\usepackage{amssymb}
\draftfalse

\journalname{Geophysical Research Letters}

\begin{document}

%
%

\title{Momentum and pressure balance of a comet ionosphere}

%
%




\authors{Hayley Williamson\affil{1}, Hans Nilsson\affil{1}, Gabriella Stenberg Wieser\affil{1}, A. I. Eriksson\affil{2}, Ingo Richter\affil{3}, Charlotte Goetz\affil{4}}


\affiliation{1}{Swedish Institute of Space Physics, Bengt Hultqvists Väg 1, SE-981 92 Kiruna, Kiruna, Sweden}
\affiliation{2}{Swedish Institute of Space Physics, Uppsala, Sweden}
\affiliation{3}{Institut f{\"u}r Geophysik und extraterrestrische Physik, TU Braunschweig, Braunschweig, Germany}
\affiliation{4}{ESA/ESTEC, Keplerlaan 1, 2201A2, Noordwijk, Netherlands}




\correspondingauthor{Hayley Williamson}{hayley@irf.se}




\begin{keypoints}
\item Momentum flux of Rosetta ion data shows the evolution of a cometary ionosphere, including boundary regions
\item Magnetic pressure is on the order of the total ion momentum flux and roughly corresponds with cometary ion momentum flux
\item The cometary pickup ions in the solar wind ion cavity take over the role of the solar wind ions outside the cavity
\end{keypoints}

%
%

%
%


\begin{abstract} 
We calculate the momentum flux and pressure of ions measured by the Ion Composition Analyzer (ICA) on the Rosetta mission at comet 67P/Churyumov-Gerasimenko. The total momentum flux stays roughly constant over the mission, but the contributions of different ion populations change depending on heliocentric distance. The magnetic pressure, calculated from Rosetta magnetometer measurements, roughly corresponds with the cometary ion momentum flux. When the spacecraft enters the solar wind ion cavity, the solar wind fluxes drop drastically, while the cometary momentum flux becomes roughly ten times the solar wind fluxes outside of the ion cavity, indicating that pickup ions behave similarly to the solar wind ions in this region. We use electron density from the Langmuir probe to calculate the electron pressure, which is particularly important close to the comet nucleus where flow changes from antisunward to radially outward.
\end{abstract}

\section*{Plain Language Summary}
To understand the atmosphere of a comet, we must understand how its charged particles interact with the solar wind. Here we look at momentum flux, the amount of momentum carried by particles flowing through a certain area, measured at the comet 67P/Churyumov-Gerasimenko by the Rosetta spacecraft. We find that the total momentum flux changes the amount we would expect from the comet getting closer to the Sun. When in the part of the atmosphere with no solar wind, the ions coming from the comet have more momentum flux than the solar wind ions elsewhere in the comet atmosphere. The increase matches what we expect from the comet getting closer to the Sun, which increases the total density of the comet atmosphere. The cometary ions replace the solar wind ions in the atmosphere, so the total stays the same. The magnetic field varies with the cometary ions. We find electron pressure with measured electron data. It is higher than the momentum flux, especially in the part of the atmosphere close to the comet nucleus. This means that the atmosphere here is primarily moving outward from the comet, while farther away from the nucleus it is mainly flowing away from the Sun.

%
%

\section{Introduction} \label{section:Intro}
A cometary atmosphere provides an opportunity to study the interaction of an escaping ionosphere with the solar wind. The ionosphere is born when neutral particles, typically volatiles such as H$_2$O and CO$_2$, are outgassed from the comet surface, forming a coma \cite{Nilsson2015}. The coma is then ionized via photoionization, charge exchange, or impact ionization \cite{Galand2016,Wedlund2017,Heritier2018}. The density of the ionosphere is dependent on comet activity, which increases with decreasing heliocentric distance. This creates a highly dynamic environment with a variety of plasma interaction phenomena, particularly when ions of cometary origin encounter the convective electric field carried by the solar wind. The Rosetta spacecraft orbited the nucleus of comet 67P/ Churyumov-Gerasimenko (67P) from August 2014 to October 2016 and observed the evolution of a cometary ionosphere and interaction with the solar wind through a range of heliocentric distances and activity. Previous cometary missions involved distant flybys, while Rosetta's trajectory stayed within a few hundred km of the nucleus, often down to few tens of km. Rosetta's distance from the comet nucleus did not vary much over the course of its mission, with the exception of two ``excursions", one on the dayside and one on the nightside of the comet. Comet activity increased as 67P drew closer to the sun, expanding the ionosphere and changing the environment Rosetta observed. In general, however, comet 67P best exemplified solar wind interactions typical of a weak comet \cite{Bagdonat2002,Wedlund2017}.

The plasma forming the ionosphere at as weak comet such as 67P initially consists of cold ions moving radially outward and fast photo-electrons with an energy of several eV. These ions have the velocity of the parent neutral gas, about 0.5 - 1 km/s. The faster electrons drag the slower ions with them through ambipolar diffusion, i.e. when the electrons' lighter mass gives them a larger thermal velocity than the ions, separating the two and creating an electric field. This electric field is believed to be a source of acceleration for the ions \cite{Vigren2015,Vigren2017a,Bercic2018,Odelstad2018,Nilsson2018}. Farther away from the comet nucleus, the cometary ions can undergo a ``pick-up" process, where they are accelerated by the solar wind electric field. They then begin to gyrate around the magnetic field lines. When the magnetic field is weak, they can escape from the comet before making a full gyration \cite{Coates2004}. These acceleration processes mean that there are two distinct cometary ion populations: the weakly accelerated cold ions with a significant radial velocity component and the energetic pickup ions. Newborn ions have a large gyro radius and thus are initially accelerated along the electric field direction. For the ions to gain this velocity, they undergo a momentum transfer with the solar wind, where both pickup and solar wind ions are deflected and the solar wind is slowed down. This is known as ``mass loading" and was observed by Rosetta \cite{Broiles2015,Behar2016,Behar2016a,Volwerk2016,Glassmeier2017}. At comet 67P, the cold cometary ions have energies on the order of a few to a few ten eV, while the pickup ions have energies from around 100 eV to a few keV, increasing near perihelion \cite{Nilsson2015a}. In addition to the cometary ions are the energetic solar wind ions, composed primarily of H$^+$, He$^+$, and He$^{2+}$. The energy of the solar wind ions varies by species, with a typical proton energy of $\sim 1$ keV.

For a sufficient outgassing rate, the solar wind ions will begin to undergo charge exchange with the cometary ions at a boundary called the cometopause \cite{Galeev1988,Cravens1989}. Closer to the comet nucleus, where the solar wind magnetic field cannot penetrate into the ionosphere, a diagmagnetic cavity will also form. At comet 67P evidence of such a diamagnetic cavity was seen several hundred times \cite{Goetz2016,Henri2017}. The Rosetta Plasma Consortium (RPC) also saw a period of several months inside roughly 2 AU when solar wind ions were excluded from the ionosphere up to at least 1000 km from the comet nucleus, known as the solar wind ion cavity \cite{Behar2016a,Mandt2016,Nilsson2017}. These different dynamical regimes could be seen because Rosetta observed comet 67P for a large range of heliocentric distances. Hybrid interaction models show that a denser ionosphere, caused by increasing ionization, broadens the interaction region \cite{Wedlund2017}. Thus, while Rosetta's position relative to the nucleus did not change much, the changing size of the ionosphere allowed for observations of a variety of configurations and boundaries, unlike what was previously possible with comet flyby missions.

Mass loading is fundamental for understanding cometary physics; when the solar wind interacts with the comet, the ion pickup process dictates momentum must be transferred from the solar wind to the cometary ions. The slowing down of the solar wind also means the magnetic field becomes draped and pile up, increasing in magnitude \cite{Volwerk2016,Goetz2017}. Thus momentum transfer is crucial for long-term ionospheric evolution. Calculating momentum flux allows for study of these effects, particularly for transition regions like the solar wind ion cavity. 

If the ionosphere is approximated as a one-dimensional plasma system momentum flux should be a conserved quantity. Given that the flow direction of the solar wind and the cometary ions is predominately antisunward \cite{Nilsson2015a,Nilsson2017}, the 1D magnetohydrodynamics continuity equation should hold true:
\begin{equation} \label{eq:cont}
    \frac{\partial}{\partial x} \left(  \rho u^2 + P + \frac{B^2}{2\mu_0}   \right) = 0
\end{equation}
Here the first term is the dynamic pressure, dependent on the bulk velocity, the second term is the scalar plasma pressure (i.e. the sum of ion and electron pressure), and the third term is the magnetic pressure where $\mu_0$ is the vacuum permeability. The sum of the dynamic and scalar pressure is equivalent to the momentum flux, i.e. the amount of momentum carried by particles per unit area per unit time. We can compare the momentum flux and magnetic pressure calculated from Rosetta data, as the sum of the two should be conserved for antisunward flow. Close to the comet we expect an outward radial flow of cold plasma, meaning a 1D ionospheric model is no longer appropriate. For this reason we also examine the electron thermal pressure at an electron temperature of 10 eV, considered to be indicative of recently ionized plasma \cite{Eriksson2017}. Recently ionized plasma should still be expanding outward from the comet nucleus like the neutral gas \cite{Edberg2015,Heritier2017,Vigren2017a,Odelstad2018}, so the electron thermal pressure should give an estimate of the contribution of radial flow to the total momentum flux. Equation \ref{eq:cont} dictates that the sum of the cometary ion antisunward momentum flux, solar wind momentum flux, and magnetic pressure should remain roughly constant so that momentum is conserved in the x direction.

\section{Instrument Description}

The data used in this study comes from the Ion Composition Analyzer (ICA), Langmuir probes (LAP), Mutual Impedance Probe (MIP) and magnetometer (MAG) in the RPC. ICA is an ion mass spectrometer able to distinguish between ion masses of 1, 2, 4, 8, 16, and 32 amu per elementary charge \cite{nilsson2007}. The instrument is capable of measuring ions in the energy range of a few eV to 40 keV per charge for positively charged ions and has an energy resolution of approximately 7\%. The instrument field of view is divided into 16 azimuthal anode sectors, each with a field of view of $22.5^{\circ} \times 4^{\circ}$. The instrument has an additional 16 elevation angles ranging $\pm 45^{\circ}$ from the central viewing plane. Due to spacecraft obstructions, the total field of view is approximately $2 \pi$ sr. A complete scan through all elevation angles and energies is done every 192 s. The data we used as input for our moment calculations is in the ESA Planetary Science Archive (PSA) as L4 PHYS MASS, containing differential flux for each species.

The plasma density measurements used derive from the LAP and MIP instruments. LAP obtains electron number density, spacecraft potential, and electron temperature from two spherical Langmuir probes \cite{Eriksson2007,Eriksson2017}. The probes are mounted on booms protruding from the spacecraft, but ion and electron measurements can still be affected by the spacecraft potential, found to be $\sim -5 - -10$V throughout the mission \cite{Odelstad2017}.  An independent electron density measurement is obtained by MIP \cite{Trotignon2007,Henri2017}. Compared to LAP, MIP densities typically have higher accuracy but less time resolution, less dynamic range, and coarser quantization. Several data products combining the data from these two instruments await publication in PSA. We here use the low time resolution data set known as NED, available from AMDA \cite{Jacquey2010}. This data set uses running calibrations of the LAP spacecraft potential, available throughout the mission, to MIP density values, thus combining large dynamic range and wide time coverage with good accuracy.

Finally, we use data from MAG, made of two triaxial fluxgate magnetometers mounted on a 1.5 m boom. MAG has a maximum magnetic field vector sample rate of 20 vectors per second with a resolution of 31 pT. Because the spacecraft had a high level of magnetic noise pollution, uncertainties of the magnetic field measurements can be up to the order of a few nT. \cite{Glassmeier2007,Goetz2016,Richter2019}. 

\section{Methods}

\subsection{Ion moment calculations}

To determine the momentum flux throughout the duration of the Rosetta mission, we calculate the second order ion moment, the momentum flux tensor, which after subtraction of the bulk flow momentum flux components gives the pressure tensor. Each particle species detected by ICA can be represented using a distribution function in phase space: $f(\mathbf{v})$, where $\mathbf{v} = (v_x,v_y,v_z)$ represents the particle's velocity. The distribution function can be written in terms of the measured differential number flux and energy:
\[f(\mathbf{v}) = \frac{m^2}{2E} \, j(E, \Omega)\]
where m is the mass of the particle, E is the particle energy, and the differential number flux j is a function of energy and solid angle $\Omega$ \cite{Franz2006,nilsson2007}. For the purposes of this paper, the ions detected by ICA are split into solar wind ions (H$^+$, He$^{2+}$, and He$^+$) and cometary ions. These heavy cometary ions are assumed to be largely water ions with a mass of 18 amu \cite{Nilsson2015a}.

The ion moments for each species can be described using a tensor of order k:
\[\mathbf{M}^k = \int_{\mathbf{v}} \, \mathbf{v}^k \ f(\mathbf{v}) \ d^3 \mathbf{v} \]
where $k = 0$ gives number density, $k = 1$ gives velocity, and $k = 2$ gives the momentum flux tensor used here. Number density and velocity have been calculated for the Rosetta mission in \citeA{Nilsson2017}. The momentum flux components below are the diagonal elements of the tensor, as the off-diagonal elements are typically negligible. Thermal pressure is then the momentum flux minus the dynamic pressure \cite{Franz2006}. To transform the flux expressions from the instrument angular coordinates to the Cartesian cometocentric solar equatorial (CSEQ) reference frame, we need the sines and cosines of the angles defining instrument pointing ($\theta$ and $\phi$) with respect to the CSEQ basis vectors as shown below. In CSEQ coordinates x points towards the Sun, z is the component of the Sun's north pole
orthogonal to the x axis, and y completes the right handed system.
\begin{eqnarray}
    \Phi_{Pxx} =& m \sum_{E,\phi,\theta} j v_x \cos^2 \theta \cos \phi \ \Delta E \Delta \phi \Delta \theta \\
    \Phi_{Pyy} =& m \sum_{E,\phi,\theta} j v_y \cos^2 \theta \sin \phi \ \Delta E \Delta \phi \Delta \theta \\
    \Phi_{Pzz} =& m \sum_{E,\phi,\theta} j v_z \cos \theta \sin \theta \ \Delta E \Delta \phi \Delta \theta
\end{eqnarray}

\begin{eqnarray} \label{eq:pres}
    P_{x} =& \Phi_{Pxx} - m u_x^2 n \\
    P_{y} =& \Phi_{Pyy} - m u_y^2 n \\
    P_{z} =& \Phi_{Pzz} - m u_z^2 n
\end{eqnarray}

Here $\Phi_{P_{ii}}$ represent the diagonal components of the momentum flux tensor and $P_i$ the diagonal components of the ion pressure tensor. The total thermal pressure $P$ is the mean of the three $P_i$ values and $u_i$ is the bulk flow velocity, i.e. the first order moment. The magnitude of the momentum flux can be compared to the magnetic pressure from \ref{eq:cont} using the magnetic field magnitude from MAG. This allows us to compare where in the comet ionosphere the various ion groups, i.e. the momentum flux of the cometary ions $\Phi_{PC}$ and solar wind momentum flux $\Phi_{PSW}$, as well as the in situ magnetic field, carry the majority of the momentum.

We also calculate the momentum flux of the upstream solar wind for comparison to the ICA data. To do so, we first obtain the density and velocity of the undisturbed solar wind near 1 AU from NASA/GSFC's OMNI data set through OMNIWeb, taken by the Advanced Composition Explorer (ACE) and Wind spacecraft. The density of the solar wind given by OMNI is scaled by $1/r^2$ and both density and velocity are time shifted to the location of comet 67P \cite{Vennerstrom2003,Opitz2009,Edberg2010,BeharThesis}. The solar wind ion flow is supersonic so its momentum flux is almost entirely due to the ion bulk flow: 
\begin{equation}
\Phi_{Psw} = n_i m_i v_i^2
\end{equation}

From equation \ref{eq:pres}, $\Phi_{Pxx}$ will be the sum of the dynamic pressure and the ion thermal pressure $\Phi_{Pxx} = P_{dyn} + P_{ion}$, so we can rewrite equation \ref{eq:cont} as 
\begin{equation}
    \frac{\partial}{\partial x} \left(\Phi_{Pxx} + P_e + \frac{B^2}{2\mu_0}\right) = 0
\end{equation}
where $P_e$ is the electron pressure such that $P = P_{ion} + P_e$ and $P_{ion}$ has been included in the momentum flux term. However, $P_e$ cannot be absorbed into the momentum flux term as ICA only detects ions, and so it must be found separately. Thus our 1D MHD approximation also requires knowing the electron pressure.

\subsection{Lower energy plasma}
While ICA can, in principle, detect ions down to 0 eV due to low spacecraft potentials of $< -5$ V for most of the mission \cite{Odelstad2017}, determining the moments of these cold ions is difficult. Despite good observations, angular coverage for these low energies is limited. The geometric factor of the instrument additionally is different at very low energies. Modeling of the spacecraft and instrument shows that distortion of the instrument FOV and sensitivity are dependent on spacecraft potential \cite{Bergman2020,Bergman2020b}. Due to these effects, we do not use ICA data here to estimate momentum flux of ions with energies of only a few eV. However, estimating the contribution of this cold plasma to the dynamics of the ionosphere is necessary, as it is estimated to have a density of roughly $10^2 - 10^3 \ \mathrm{cm^{-3}}$, about two orders of magnitude higher than the average accelerated water ion number density \cite{Nilsson2017,Nilsson2017b}. Instead we find thermal pressure from the LAP and MIP instruments, for an assumed $T_e = 10$ eV \cite{Eriksson2017}. This temperature is consistent with photoelectron temperatures found in both the RPC Ion and Electron Sensor (IES) data and ionospheric modeling \cite{Galand2016,Madanian2016}. Since $T_e$ is about an order of magnitude larger than $T_i$, we only calculate the electron thermal pressure, as the sum of the electron and ion thermal pressures would be expected to be roughly equal to the electron thermal pressure. We expect the electron pressure to be important primarily close to the nucleus, since as stated above it represents weakly accelerated plasma moving radially outward.

\section{Results}

\begin{figure}
\centering
\noindent\includegraphics[trim={0.7cm 2cm 0.7cm 2cm},clip,width=1.1\textwidth] {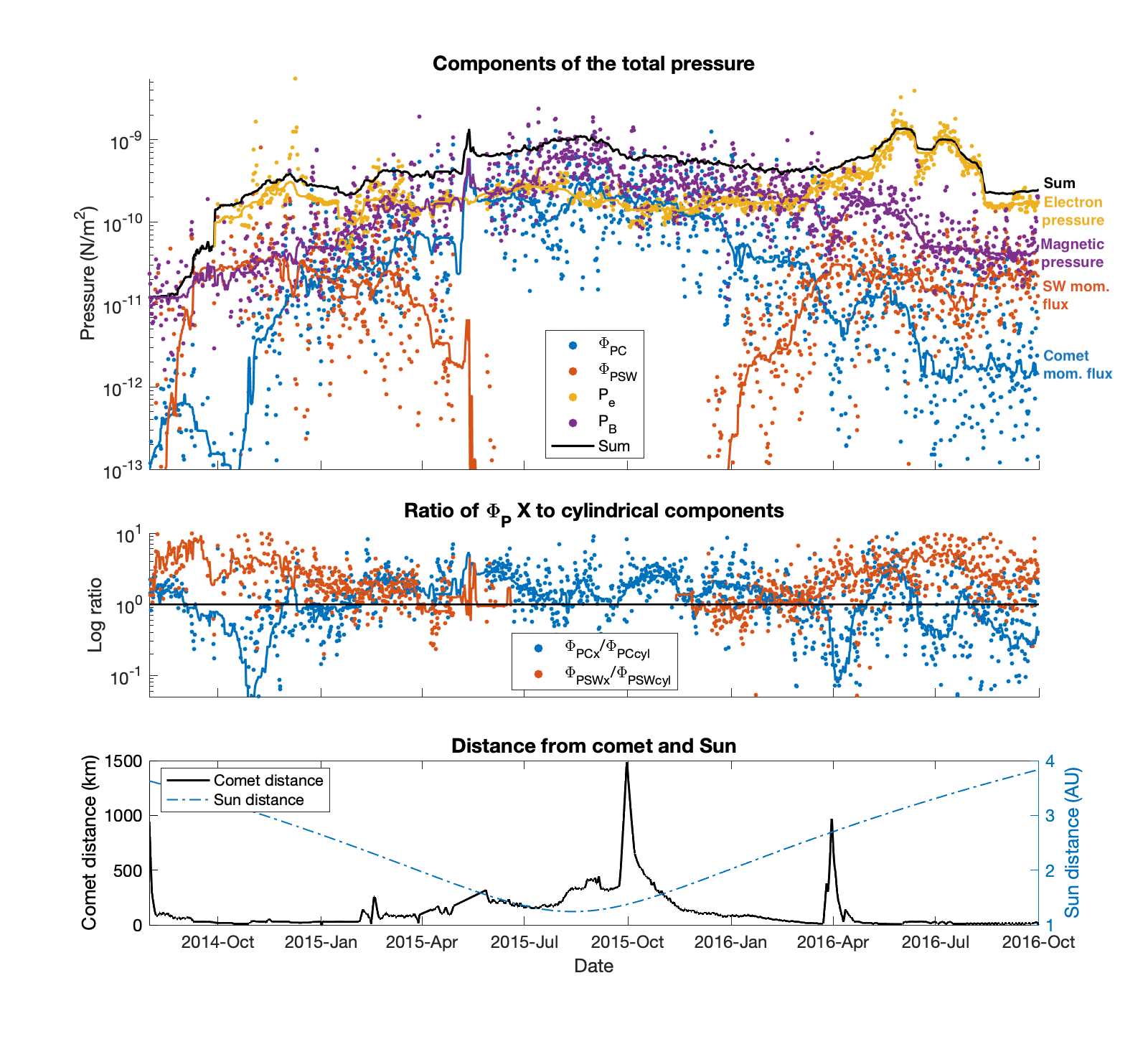}
\caption{Contributions to the total pressure for the time period of Rosetta orbiting the comet nucleus. In panel (a) red indicates solar wind ions, blue cometary ions, yellow electron pressure, and purple magnetic pressure, with solid lines indicating a running median value. The black solid line is a median value for the sum of all the components. All are in units of N/m${^2}$. Panel (b) shows the ratio of the x components of the cometary (blue) and solar wind (red) momentum flux to the cylindrical components in CSEQ coordinates with a line for the running median. Panel (c) shows the spacecraft distance from the comet nucleus, with the two excursions visible in October 2015 and April 2016. The blue dashed-dotted line is the heliocentric distance.}
\label{fig:momentum}
\end{figure}

Figure \ref{fig:momentum}(a) shows the ICA ion momentum flux magnitude for cometary and solar wind ions, electron thermal pressure at 10 eV, and magnetic pressure for the entirety of the Rosetta spacecraft orbiting the nucleus of 67P. Panel (c) shows radial distance of the spacecraft from the comet in km and the heliocentric distance in AU. The full dataset is quite noisy, making trends difficult to see. In order to better understand the long-term behavior of the ions, we therefore averaged the data every 12 hours, roughly the rotation period of the comet nucleus. This allows us to smooth the data significantly, as well as removing trends related to features on the surface of the comet, which can vary in their outgassing rates \cite{Lauter2018} and are not of interest here. However, there is still significant scatter even after averaging over every 12 hours, so we include a line showing the running median value with a 30 day window to indicate trends on the scale of cometary activity evolution. The median was chosen over the mean due to the wide scatter and number of outliers. These two methods of smoothing the data allow us to examine the changes in total momentum flux relative to heliocentric distance. The energy table for ICA was changed at the end of 2014, so previous energies for cometary ions had an uncertainty of about 10 eV \cite{Nilsson2015a}. The instrument was also run sparsely in the beginning of the mission, giving poor statistics. Therefore the values before the end of 2014 have larger uncertainties.

The most prominent feature in Figure \ref{fig:momentum} is the solar wind ion cavity from mid-April 2015 to December 2015. Comet activity increased enough near perihelion for the cometary atmosphere to present a significant obstacle to the solar wind in the region where the spacecraft orbited. As a result, only cometary ions were detected \cite{Behar2017}. The transition between the region of both solar wind and cometary ion detections to solely cometary ion detections is clearly visible, with momentum flux rapidly decreasing for the solar wind ions and increasing for the cometary ions. While the ICA cometary ion energy is lower than the energy of the solar wind outside the cavity, the increased density of the cometary ions in the cavity leads to the cometary ion momentum flux exceeding the solar wind momentum flux seen before April 2015 and after January 2016. When the momentum flux is scaled according to the square of heliocentric distance from the comet, the cometary momentum flux at perihelion differs from the solar wind momentum flux outside the solar wind ion cavity by less than an order of magnitude, indicating this increase is largely due to higher comet activity near perihelion. The various observations of the diamagnetic cavity, where the magnetic field is not observed, are not visible in this data set due to the data averaging, which eliminates small-scale features.

The electron pressure is often the largest contributor to the total, with the exception of in the solar wind ion cavity. In the solar wind ion cavity, when the cometary ion density is the highest, the electron pressure is roughly equal to the cometary ion momentum flux. Towards the end of the mission, when the spacecraft is closest to the nucleus, the electron pressure becomes nearly two orders of magnitude larger than all other contributors to the total momentum flux, including magnetic pressure. The magnetic pressure seen in Figure \ref{fig:momentum}(a) typically follows the cometary ion or solar wind momentum flux rather than the electron pressure. The increase of the magnetic pressure in the solar wind ion cavity occurs when the magnetic field strength increases with mass loading and deflection of the solar wind. Solar wind deflection increases with decreasing heliocentric distance \cite{Behar2016a}, causing a subsequent increase in magnetic pressure. Interestingly, the magnetic pressure varies with the momentum flux on shorter timescales, including in the solar wind ion cavity where it varies with the cometary ions (e.g. April-July 2016). Changes in the magnetic field corresponding to changes in ion velocity during the dayside excursion have been noted previously in \citeA{Volwerk2019a}, but no conclusions were drawn on the cause of this covariance. 

As the direction of the momentum flux is important for mass loading, Figure \ref{fig:momentum}(b) shows the ratio of momentum flux for solar wind and cometary ions split into CSEQ components x and cylindrical, where $\Phi_{Pcyl} = \sqrt{\Phi_{Pyy}^2 + \Phi_{Pzz}^2}$. The x component of the momentum flux for both solar wind and cometary ions is directed antisunward (negative x) for the vast majority of the mission. Here we show the log of the absolute value of the ratio, since the ratio is almost always negative with very few exceptions. The ratio in panel (b) shows that the solar wind is almost always primarily in the negative x direction, i.e. the absolute value of the ratio is greater than one. For the cometary ions, there is more variation. For larger heliocentric distances and thus less cometary activity, the cylindrical component can be larger than the x component. There is a particularly large dip in the ratio when the spacecraft made its nightside excursion in April 2016. In this excursion, the spacecraft moved up to 1000 km away from the nucleus in the shadow of the comet. Panel (b) shows that in this region the cometary ions are no longer moving in an antisunward direction. This may be worthy of a future, more in-depth study. A similar decrease is seen at the end of the mission when the spacecraft is very close to the comet. This change is expected, as the ionosphere transitions to a mostly-radial flow at these small distances \cite{Heritier2017,Bercic2018}. However, the ratio is greater than one for the majority of the solar wind ion cavity, showing that the cometary ion momentum flux in this part of the ionosphere is most similar to that of the solar wind.

\begin{figure}
\centering
\noindent\includegraphics[trim={0.7cm 2cm 0.7cm 2cm},clip,width=\textwidth] {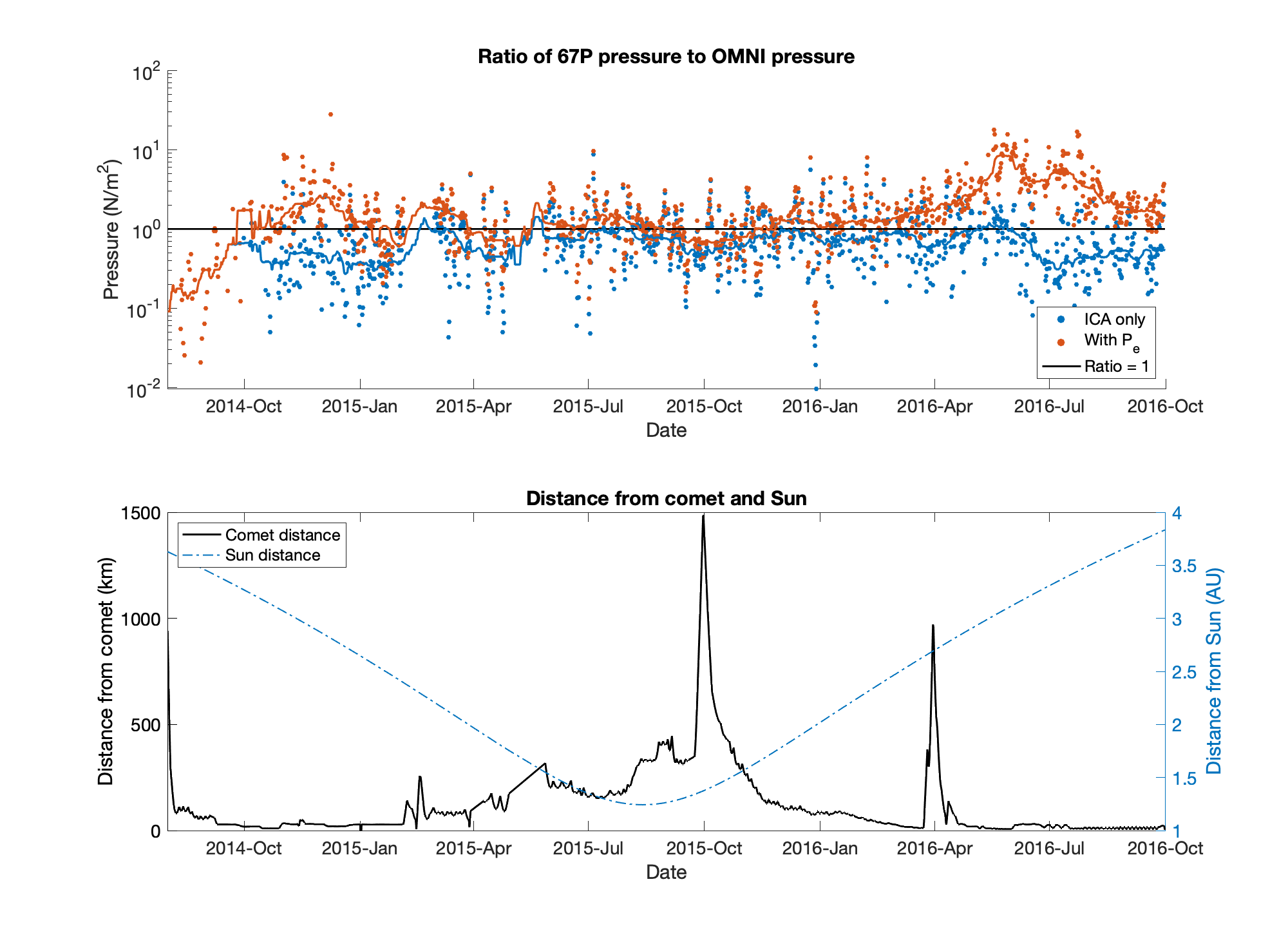}
\caption{Panel (a) shows the ratio of Rosetta momentum flux to the undisturbed solar wind momentum flux retrieved from OMNIWeb and scaled to the location of 67P. Blue shows the sum of solar wind ions, cometary ions, and magnetic pressure, while red also includes electron pressure. The solid blue and red lines are running median values and the black line is a reference of the ratio = 1. Panel (b) shows the distance of the spacecraft to the comet nucleus and the comet to the Sun.}
\label{fig:omni}
\end{figure}

In Figure \ref{fig:omni}(a) we compare the momentum flux seen by Rosetta with the solar wind momentum flux outside of the cometary ionosphere. Because the electrons seen by LAP lack a strong antisunward component \cite{Edberg2015,Eriksson2017}, we plot the ratio of the sum of the ion, magnetic, and electron pressures in red and sum of ion and magnetic pressures in blue to the OMNI momentum flux. The ratio between the Rosetta total pressure both with and without electron pressure and the undisturbed solar wind momentum flux does not change much over the mission, indicating that the change in total momentum flux seen by Rosetta is largely due to heliocentric distance. Particularly for the case without electron pressure (blue), the ratio remains close to one for the duration of the mission. The electron pressure adds additional momentum flux relative to the OMNI data at the end of the mission during the same time period when it dominates the total momentum pressure. The large amount of scatter in the data is likely because the solar wind ions at Rosetta undergoes charge exchange with the neutrals, as well as compression and rarefaction due to mass loading \cite{Behar2016a,Volwerk2016}.

\section{Discussion and Conclusions}
The data show that the momentum flux changes consistently with heliocentric distance for the portion of the comet ionosphere seen by Rosetta, with magnitudes of $\mathrm{\sim 10^{-10}-10^{-9} \ N/m^2}$ with an increase near comet periapsis, even through varying rates of comet activity and solar wind conditions. This implies that the pickup cometary ions take up the bulk of the antisunward momentum flux in the ionosphere in the solar wind ion cavity, where high densities lead to an increase in the momentum flux, as would be expected in a mass loading scenario. While solar wind ions are absent in the solar wind ion cavity, the magnetic field is still present and shows a clear contribution to the total momentum flux, in addition to varying with the cometary ions. On either side of the solar wind ion cavity, seen in early 2015 and early 2016, the momentum flux of the solar wind and cometary ions are roughly equal, unlike at the beginning and end of the mission, where the solar wind ion momentum flux dominates over the pickup ions. It is known that Rosetta observed multiple boundary regions throughout its mission \cite{Mandt2016}; the data here shows evidence of the ionosphere transitioning from being solar wind origin ion dominant to cometary origin ion dominant near perihelion. 

However, while there is a transition region, the roughly constant total momentum flux indicates that the solar wind ion cavity is not a significant boundary in terms of energy input into the ionosphere system. Thus while the solar wind ion cavity plays an important role in the relative contributions of the different ion populations, it does not significantly change the fundamental momentum budget. It is possible that this is because the solar wind ion cavity is not a barrier to the solar wind magnetic field or electrons, which in fact shows an increase in magnetic pressure (see purple line in Figure \ref{fig:momentum}(a)) much like the cometary ion momentum flux. This illustrates Equation \ref{eq:cont}, showing that the momentum flux does indeed remain roughly constant.

We also conclude that the pickup ions (i.e. those seen by ICA) essentially replace the solar wind ions in the solar wind ion cavity. Figure \ref{fig:momentum}(b) shows that the pickup ions' flux vector components in the ion cavity are similar to the solar wind, with their antisunward momentum flux becoming greater than that of the solar wind ions before and after the solar wind ion cavity. Therefore these ions have been accelerated so that they now move similarly to solar wind ions, rather than radially away from the nucleus. Additionally, the ratio of the Rosetta momentum flux to OMNI momentum flux seen in Figure \ref{fig:omni} remains roughly constant, with a median value around 1, showing that the cometary ions are able to maintain a consistent level of momentum flux even without the presence of solar wind ions. This is as expected, as the presence of a region where the solar wind is excluded implies a high enough pressure, or momentum flux, to create a balance. The pickup ions behaving similarly to solar wind ions is further supported by \citeA{Masunaga2019}, which shows evidence of cometary ions both outside and inside the diamagnetic cavity having a flow direction consistent with deflection, i.e. opposite to the direction of the solar wind electric field, indicating mass loading.

An additional transition region is seen after the solar wind ion cavity, where the electron pressure dominates the total momentum flux, including the magnetic pressure, by almost two orders of magnitude. These high electron pressure values occur when Rosetta is close to the nucleus (Figure \ref{fig:momentum}). Here we do not expect local momentum balance to the upstream solar wind as in equation \ref{eq:cont}, as we are inside an expanding and escaping ionospheric plasma. Furthermore, we may note that our assumed electron temperature of 10 eV may sometimes be an overestimate, as a large fraction of the electrons have been cooled to $\lesssim 0.1$ eV \cite{Eriksson2017,Gilet2017a}. While there is a suprathermal electron population seen by IES, this is evidently much lower density and so likely does not change the average temperature much for this approximation \cite{Madanian2016}. Much of the apparent $P_e$ increase in summer 2016 could be due to such an overestimation of $T_e$. It is clear that the electron pressure merits a dedicated investigation outside the scope of this paper, possibly also including higher energy electrons, which can contribute significantly to total pressure even if their density is low \cite{Mandt2016}.

In conclusion, studying momentum flux gives insight into the overall dynamics of the comet 67P ionosphere as measured by Rosetta due to its importance in mass loading and ion pickup. We see that generally the solar wind ions carry the bulk of the antisunward momentum, while radially flowing cold plasma contribute the most to the total. In the solar wind ion cavity, a transition occurs, where magnetic pressure increases and cometary pickup ions have momentum flux similar to the mass-loaded solar wind in heliocentric distance scaled magnitude. The direction of the cometary momentum flux in the solar wind ion cavity is antisunward, also like that of the solar wind. Thus in the solar wind ion cavity, the mass loaded pickup ions behave like the solar wind ions observed at lower levels of comet activity. The magnetic pressure and cometary ion momentum flux appear to covary. Despite this transition, the total pressure of the portion of the ionosphere Rosetta observed appears to vary mainly with heliocentric distance, so this boundary is not a significant barrier for momentum input.  In the latter half of the mission, LAP data shows an electron pressure multiple orders of magnitude larger than all other components to the total pressure. Therefore close to the comet nucleus the mass-loaded solar wind, pickup ions, and magnetic field are less important for ionospheric dynamics than the cold plasma flowing in the direction of neutral outgassing. Because of its long term mission, Rosetta was able to observe various dynamical regimes in the ionosphere of comet 67P, and this is reflected in the changing contributions of various components to the total pressure balance. As predicted by the magnetohydrodynamics continuity equation, the sum of the pressure and momentum flux components varies only with comet activity, not region of the ionosphere, indicating a conserved quantity.

\acknowledgments
Funding support comes from Swedish Research Council contract 2015-04187. We acknowledge use of NASA/GSFC's Space Physics Data Facility's OMNIWeb service for OMNI data and the Planetary Science Archive (PSA, http://archives.esac.esa.int/psa)) for Rosetta MAG, LAP, MIP, and ICA data. We also acknowledge the use of AMDA (amda.cdpp.eu) for LAP and MIP data. Data used in this study will be available through the Swedish National Data Service. C. Goetz is supported by an ESA Research Fellowship.


%
%

\bibliography{Rosetta}

%

\end{document}